\documentclass[aps,pra,10pt,amssymb,amsfonts,floatfix]{revtex4-2}
\usepackage{bm}
\usepackage{dcolumn}
\usepackage{bm}
\usepackage{mathbbol}
\usepackage{graphicx}
\usepackage{epstopdf} 
\usepackage{amsmath,amsthm}
\usepackage{times}
\usepackage{lipsum}
\usepackage{nicefrac}
\usepackage{amsfonts}
\usepackage{amssymb}
\usepackage{amsmath} 
\usepackage{subfigure}
\usepackage{multirow} 
\usepackage{tabularx} 
\usepackage{array}
\usepackage{float}
\usepackage{units}
\usepackage{hyperref}
\usepackage{xcolor}

\begin{document}

\title{Generalized Survival Probability}

\author{David A. Zarate-Herrada$^{1}$, Lea F. Santos$^{2}$ and E. Jonathan Torres-Herrera$^{1}$}

\affiliation{$^1$Instituto de F{\'i}sica, Universidad Aut\'onoma de Puebla, Apt. Postal J-48, Puebla, Puebla, 72570, Mexico}
\affiliation{$^2$Department of Physics, University of Connecticut, Storrs, Connecticut 06269, USA}

\begin{abstract}
The survival probability measures the probability that a system taken out of equilibrium has not yet moved out from its initial state. Inspired by the generalized entropies used to analyze nonergodic states, we introduce a generalized version of the survival probability, and discuss how it can assist studies of the structure of the eigenstates and ergodicity.
\end{abstract}

\maketitle
\section{Introduction}

The square overlap between a given initial state $|\Psi(0)\rangle $ and its time-evolved counterpart $|\Psi(t)\rangle $, 
\begin{equation}
SP (t) = |\langle  \Psi(0) | \Psi(t)\rangle|^2,
\label{Eq:SP1}
\end{equation}
gives the probability for finding the system still in its initial state at time $t$. This quantity is known as survival probability, return probability, or simply the fidelity between the initial and the evolved state. This quantity has been extensively investigated since the early decades of quantum mechanics, initially in the context of the uncertainty relation between time an energy~\cite{Krylov1947,Fock1947,Fock1962}. As stated by Fock in Ref.~\cite{Fock1962},
\begin{quote}
[the time-energy uncertainty relation] may be viewed as
a consequence of the general theorem of Fock and
Krylov on the connection between the decay law
and the energy distribution function.
\end{quote}
The ``connection'' stated in the quote above refers to the fact that the survival probability (decay law) is the absolute square of the Fourier transform of the energy distribution of the initial state (energy distribution function). That is, for a state evolving according to a Hamiltonian $H$, whose eigenvalues and eigenstates are given by $E_\alpha$ and $|\alpha \rangle$, one has that $|\Psi(t) \rangle = \sum_{\alpha} C_{\alpha}^{(0)} e^{-iE_{\alpha} t} |\alpha\rangle$ and the survival probability can be written as
\begin{equation}
SP (t) = \left| \sum_{\alpha} |C_{\alpha}^{(0)}|^2 e^{-iE_{\alpha} t} \right|^2 = \left| \int \rho(E)  e^{-iE t} dE
\right|^2,
\label{Eq:SP2}
\end{equation}
where $C_{\alpha}^{(0)}=\langle \alpha |\Psi(0)\rangle$ and
\begin{equation}
 \rho(E) = \sum_{\alpha} |C_{\alpha}^{(0)}|^2 \delta(E - E_{\alpha})
\end{equation}
is the energy distribution of the initial state. This distribution is also known as local density of states (LDoS) or strength function, and its mean and variance are~\cite{Torres2016} 
\begin{equation}
E^{(0)} = \sum_{\alpha}  |C_{\alpha}^{(0)} |^2 E_{\alpha} \hspace{0.3 cm } \mbox{and} \hspace{0.3 cm }
\sigma^2 =  \sum_{\alpha}  |C_{\alpha}^{(0)} |^2 (E_{\alpha} - E^{(0)})^2 .
 \label{sigma} 
 \end{equation}

Both the survival probability and the LDoS are studied in a variety of different fields, from quantum chaos and nuclear physics to localization and quantum information science. These quantities received significant attention from  previous researchers of the Budker Institute in Novosibirsk, including those to which we dedicate the present paper, namely Professor Giulio Casati on the occasion of his 80th birthday in 2022, Professor Felix Izrailev on the occasion of his 80th birthday in 2021, and Professor Vladimir Zelevinsky on the occasion of his 85th birthday in 2022.

Despite the simplicity of Eq.~(\ref{Eq:SP1}), the evolution of the survival probability in many-body quantum systems is quite rich, with different behaviors emerging at different time scales, which reveal details about the initial state, the spectrum and the eigenstates of the considered model.
The Taylor expansion of the phase factor in Eq.~(\ref{Eq:SP2}) shows that the survival probability at very short times, $t\ll 1/\sigma$, presents a quadratic and  universal behavior, $SP(t) \approx 1 - \sigma^2 t^2$, where $\sigma$ is the width of the LDoS [see Eq.~(\ref{sigma})]. Beyond this point, but still at short times, $t\lesssim 1/\sigma$, the decay is dictated by the shape of the LDoS. The shape of $\rho(E) $ was investigated in~\cite{Casati1996,Fyodorov1996} in the context of banded random matrices, while in realistic models, the transition  from a Lorentzian to a Gaussian form with the increase of the perturbation strength was discussed, for example,  in~\cite{Bertulani1994, lewenkopf1994, frazier1996,zelevinskyrep1996,flambaum1997,Flambaum2000,Flambaum2000A,Flambaum2001a,Flambaum2001b,Kota2001PRE,Chavda,KotaBook,Santos2012PRL,Santos2012PRE,Torres2014NJP}. Depending on the initial state and the model considered, skewed Gaussians and bimodal distributions can also emerge~\cite{Torres2014PRAb}. Beyond the characteristic time for the initial depletion of the initial state, $t \sim 1/\sigma$, the survival probability exhibits a power-law decay $\propto t^{-\gamma}$ with an exponent $\gamma $ that depends on the level of ergodicity of $|\Psi(0) \rangle$ and $|\alpha\rangle$. When the LDoS is filled ergodically, $\gamma $ is determined by the bounds of this energy distribution~\cite{Khalfin1958,Ersak1969,Fleming1973,Knight1977,Fonda1978,Erdelyi1956,Urbanowski2009,Tavora2016,Tavora2017}. In contrast, when $|\Psi(0) \rangle$ and $|\alpha\rangle$ are nonchaotic states, then $\gamma $ depends on the level of correlations and multifractality among the states~\cite{Ketzmerick1992,Huckestein1994,Huckestein1999,Torres2015,Torres2015BJP,Torres2017}. But this is not yet the end of the story. In chaotic systems, where the energy-level statistics is similar to those of random matrices, the survival probability does not saturate after the algebraic decay. Instead, it reaches a value that is smaller than its infinite-time average, 
\begin{equation}
\overline{SP} = \sum_{\alpha} |C_{\alpha}^{(0)}|^4 ,
 \end{equation} 
and then grows in a ramp  until it finally saturates at $\overline{SP}$. The infinite-time average is the last term in the equation below, which is obtained from Eq.~(\ref{Eq:SP2}),
\begin{equation}
SP (t) = \sum_{\alpha \neq \beta} |C_{\alpha}^{(0)}|^2 |C_{\beta}^{(0)}|^2e^{-i( E_{\alpha} - E_{\beta}) t} +\sum_{\alpha } |C_{\alpha}^{(0)}|^4 .
\label{Eq:SP3}
\end{equation}
The interval in which $SP(t) < \overline{SP}$ is known as correlation hole~\cite{leviandier1986,pique1987,guhr1990,hartmann1991,alhassid1992,lombardi1993,michaille1999,gorin2004,alhassid2006,leyvraz2013,torresannphys,torres2017Philo,torres2018,Torres2019,schiulaz2019,lerma2019,lezama2021,Santos2021}, and there have been different proposals to measure it experimentally in systems out of equilibrium (see Ref.~\cite{dagARXIV} and references therein).  The correlation hole is a dynamical manifestation of spectral correlations, and as such can be used to detect many-body quantum chaos in experiments that do not have direct access to the spectrum, such as experiments with cold atoms and ion traps.

In this work, motivated by generalized quantities like R{\'e}nyi entropies~\cite{renyi1961measures}, the inverse participation ratio~\cite{Evers2008,Wegner1980,Soukoulis1984,Atas2013,Atas2013royal,Solorzano2021} and other similar quantities~\cite{Pilatowsky2022}, that have a prominent role in studies of localization and multifractality, we introduce the generalized survival probability, $SP_q(t)$, and its corresponding generalized LDoS, $\rho_q^{(0)}(E)$ [see the definitions below in Eq.~\eqref{eq:gsp} and Eq.~\eqref{eq:gLDoS}, respectively]. We discuss how they can help with the understanding of the structure of the eigenstates. 

Using the one-dimensional (1D) disordered spin-1/2 model often employed in the analysis of many-body localization, we compare the results for the generalized survival probability in the chaotic regime and away from it, where the duration of the power-law decay of $SP_q(t)$  becomes dependent on the value of $q$. We also compare the behavior of $SP_q(t)$ for the chaotic spin model with random matrices from the Gaussian orthogonal ensemble (GOE) and, in the latter case, provide an analytical expression for the entire evolution of the generalized survival probability.

\section{Models}

We study many-body quantum systems described by the Hamiltonian
\begin{equation}
H = H_0 + V   , 
\end{equation}
where a chosen eigenstate of $H_0$ corresponds to the initial state and $V$ is a strong perturbation that takes the system far from equilibrium. We consider initial states that have energies $E^{(0)} = \langle \Psi(0)|H|\Psi(0)\rangle$ close to the middle of the spectrum. Two Hamiltonians $H$ are investigated, one is a random matrix from the GOE and the other describes a 1D disordered Heisenberg spin-1/2 model.

\subsection{Gaussian orthogonal ensemble}
The GOE is composed by real and symmetric ${\cal{D}}\times{\cal{D}}$  matrices completely filled with random entries from a Gaussian distribution with mean zero and variance given by
\begin{equation}\label{eq:HamGOE}
\left\langle H_{jk}^2\right\rangle=
\begin{cases}
    \frac{1}{2},\quad\text{for}\quad j\neq k;\\
    1,\quad\text{for}\quad j=k.
\end{cases}    
\end{equation}
We assume that the unperturbed Hamiltonian $H_0$ is the diagonal part of $H$ and $V$ is the off-diagonal part. The model is unphysical, but allows for analytical derivations that can serve as a reference for the studies of realistic chaotic many-body quantum systems.

\subsection{Disordered spin-1/2 model}
As a physical model, we consider the 1D Heisenberg spin-1/2 model with onsite disorder that has been used in studies of many-body localization~\citep{SantosEscobar2004,santos2005loc,Dukesz2009,Nandkishore2015}. The Hamiltonian is given by
\begin{equation}
\label{eq:HamH}
H = \sum_{k=1}^{L} h_{k} S_{k}^{z} + 
J \sum_{k=1}^{L} \left( S_{k}^{x} S_{k+1}^{x} + S_{k}^{y} S_{k+1}^{y} + S_{k}^{z} S_{k+1}^{z} \right) ,
\end{equation}
where $S^{x,y,z}$ are the spin-$1/2$
 operators, $L$ is the system size, $J=1$ is the coupling strength, and $h_{k}$ are independent and uniformly distributed random variables in $[-h,h]$, with $h$ being the onsite disorder strength.
  We assume periodic boundary conditions.
The system conserves the total magnetization in the $z$-direction, $\hat{S}^{z}_{\mathrm{tot}}=\sum_{k=1}^{L}\hat{S}_{k}^{z}$. Throughout our study, we work in the largest subspace with $\hat{S}^{z}_{\mathrm{tot}}=0$ leading to ${\cal{D}}=L!/(L/2)!^2$. For finite  sizes, $H$ shows level statistics comparable to the GOE random matrices when $h \sim 0.5$, while level repulsion fades away for $h>1$. We take the unperturbed Hamiltonian to consist of the terms in the $z$-direction, $H_0 = \sum_{k=1}^{L} \left( h_{k} S_{k}^{z} + 
J   S_{k}^{z} S_{k+1}^{z}\right) $, and the perturbation to be the flip-flop term, $V = J \sum_{k=1}^{L} \left( S_{k}^{x} S_{k+1}^{x} + S_{k}^{y} S_{k+1}^{y}  \right)$.

\section{Generalized survival probability}
We define the generalized survival probability as
\begin{equation}
\label{eq:gsp}
SP_q(t)=\frac{1}{{\cal{N}}_q^2}\left|\sum_{\alpha=1}^{\cal{D}}|C_\alpha^{(0)}|^qe^{-iE_\alpha t}\right|^2 =\left|\int\rho_q(E)e^{-iE_\alpha t}dE\right|^2,  
\end{equation}
where ${\cal{N}}_q$ is a normalization constant given by
\begin{equation}
{\cal{N}}_q=\sum_{\alpha=1}^{\cal{D}}|C_\alpha^{(0)}|^q, 
\end{equation}
the parameter $q \geq0$ is a positive real number, and
\begin{equation}
\label{eq:gLDoS}
\rho_q(E)=\frac{1}{{\cal{N}}_q}\sum_{\alpha=1}^{\cal{D}}|C_\alpha^{(0)}|^q\delta(E_\alpha-E)
\end{equation}
is the generalized LDoS (gLDoS) with mean and variance given, respectively, by
\begin{equation}
\label{eq:E0q}
E_{q}^{(0)}=\frac{1}{{\cal{N}}_q}\sum_{\alpha=1}^{\cal{D}}|C_\alpha^{(0)}|^qE_\alpha \quad\text{and}\quad\sigma^2_q=\frac{1}{{\cal{N}}_q}\sum_{\alpha=1}^{\cal{D}}|C_\alpha^{(0)}|^q(E_\alpha-E_{q}^{(0)})^2.
\end{equation}
The survival probability, as defined in Eq.~(\ref{Eq:SP2}), and the mean and variance given in Eq.~(\ref{sigma}),  are recovered when $q=2$. For $q=0$, Eq.~(\ref{eq:gsp}) coincides with the spectral form factor~\cite{MehtaBook}, which is a quantity used to study level statistics in the time domain. Contrary to the (generalized) survival probability, the spectral form factor is not a dynamical quantity, since it does not depend on the initial state.

If one knows the generalized LDoS, we can obtain the generalized survival probability by performing the Fourier transform in Eq.~(\ref{eq:gsp}). We therefore start our analysis by examining the shape of $\rho_q (E)$.

\subsection{Generalized LDoS}
\begin{figure}[htpb!]
    \centering
    \includegraphics[scale=0.4]{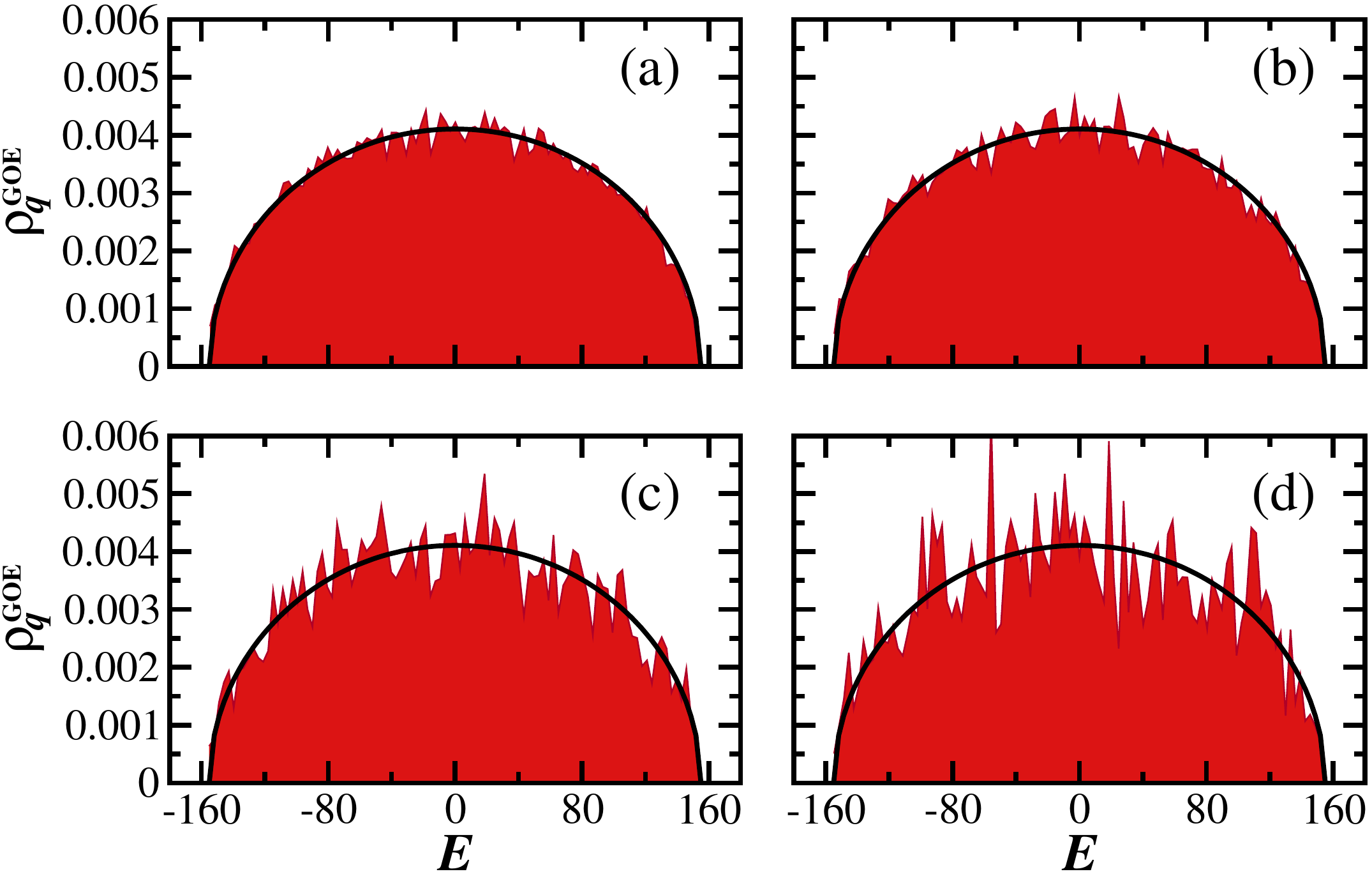}
    \caption{Generalized local density of states for GOE matrices for (a) $q=0.5$, (b) $q=1.0$, (c) $q=2.0$, and (d) $q=3.0$. Shaded areas are numerical results and the solid curves represent the semicircle law in Eq.~(\ref{eq:GOEdos}). A single disorder realization and a single initial state are considered. Matrix size is ${\cal{D}}=12,000$.
    }
    \label{fig:LDoSGOE}
\end{figure}

Figure~\ref{fig:LDoSGOE} depicts the  generalized LDoS for a single random realization of a GOE matrix and different values of $q$. We observe that the semicircular shape, typical of random matrices in the limit of large $\cal{D}$, and the length of the distribution are conserved independently of the value of $q$. This is because all eigenstates from GOE matrices are random vectors and so is the initial state, that is, $C_\alpha^{(0)}$ are random numbers from a Gaussian distribution satisfying the constraint of normalization. Even though for $q>1$, the larger components $C_\alpha^{(0)}$ get enhanced, leading to the spikes seen in Fig.~\ref{fig:LDoSGOE}~(c) and Fig.~\ref{fig:LDoSGOE}~(d), the width of the distribution is not affected by $q$. This means that after averages over random realizations, one would not notice the differences between the panels. One can then state that the robustness of the generalized LDoS for different values of $q$ is a sign of the ergodicity of the eigenstates of the system.

Since the components $|C_\alpha^{(0)}|^q$ are uncorrelated random numbers fluctuating smoothly around the average $\overline{|C_\alpha^{(0)}|^q} = {\cal{N}}_q/{\cal D}$, one sees that the gLDoS for GOE matrices coincides with the normalized density of states $P^{\text{GOE}}(E)={\cal D}^{-1} \sum_{\alpha}\delta(E_\alpha-E)$. Therefore, for GOE random matrices, we have that 
\begin{equation}
\label{eq:GOEdos}
\rho_q^{\text{GOE}}(E)= P^{\text{GOE}}(E)=\frac{1}{\pi\sigma_q}\sqrt{1-\left(\frac{E}{2\sigma_q}\right)^2},    
\end{equation}
where the standard deviation $\sigma_q=\sqrt{{\cal{D}}/2}$. In Fig.~\ref{fig:sigma}~(a), we plot $\sigma_q$ as a function of $q$ and confirm that $\sigma_q$ is indeed nearly constant for GOE.

\begin{figure}[htpb!]
    \centering
    \includegraphics[scale=0.4]{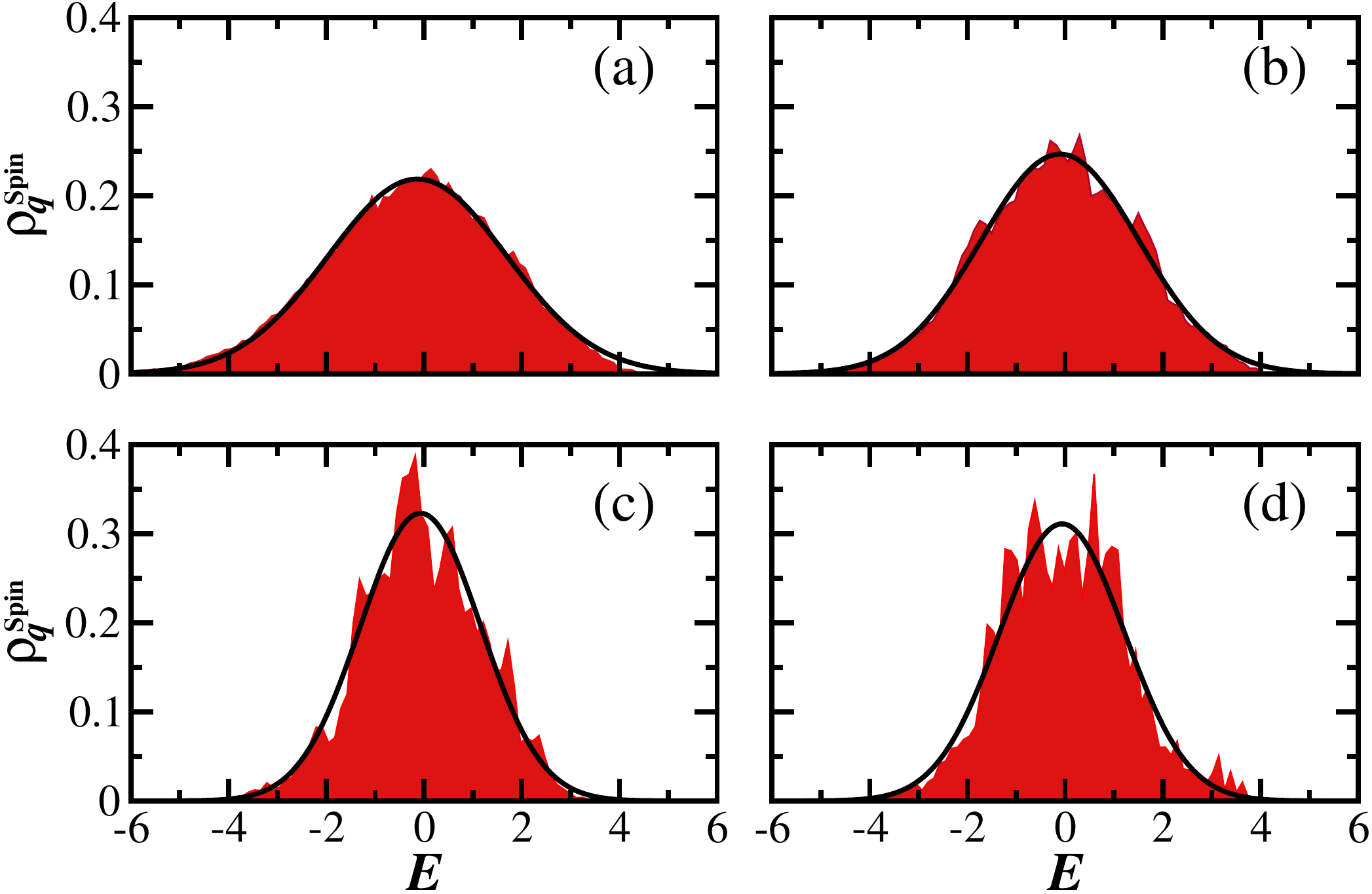}
    \caption{Generalized local density of states for the disordered spin-$1/2$ model with $h=0.5$ for (a) $q=0.5$, (b) $q=1.0$, (c) $q=2.0$ and (d) $q=3.0$. Shaded areas are numerical results and the solid curves represent the Gaussian expression in Eq.~(\ref{eq:Spindos}). A single disorder realization is considered. The system size is $L=16$ with ${\cal{D}}=12,870$.
 }
    \label{fig:LDoSspin1}
\end{figure}

\begin{figure}[htpb!]
    \centering
    \includegraphics[scale=0.5]{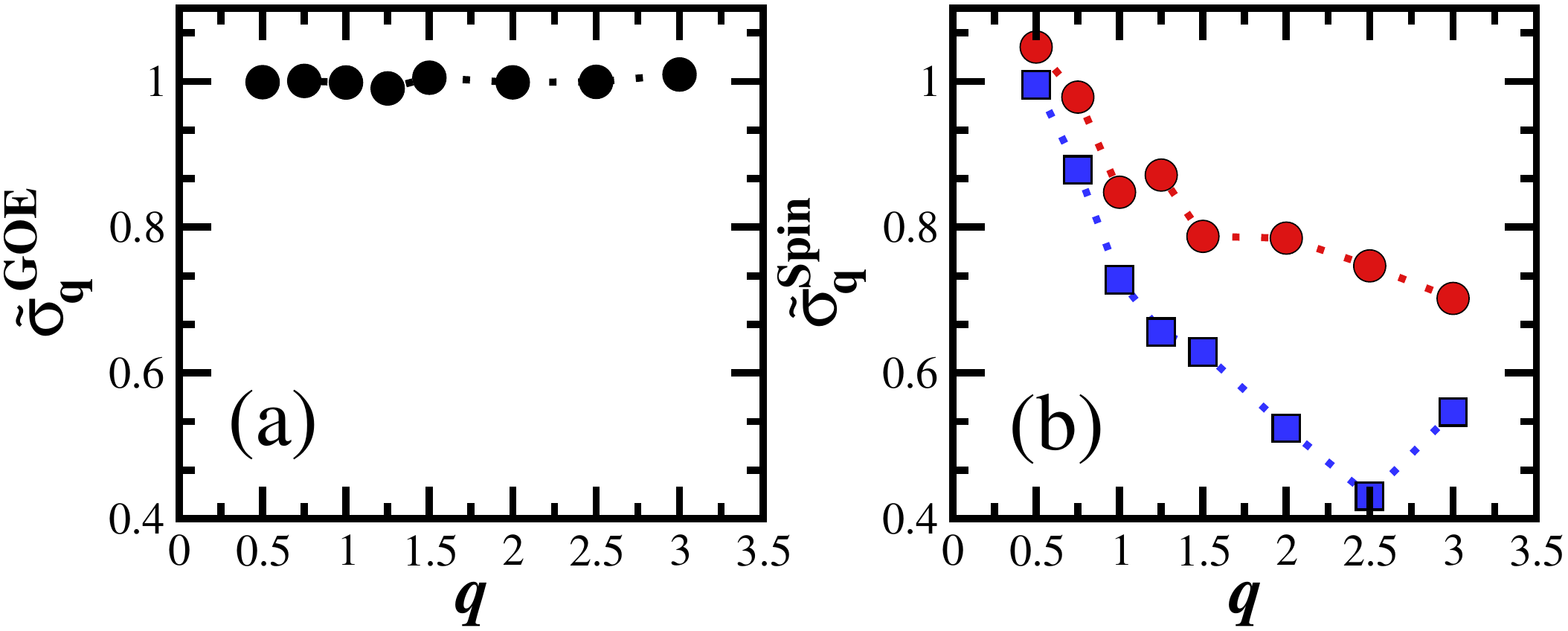}
    \caption{Width of the generalized LDoS normalized by the DoS for (a) GOE matrices, $\tilde{\sigma}^\text{GOE}_q=\sigma_q^{GOE}/\sigma_{DOS}^{GOE}$, and (b) the spin model, $\tilde{\sigma}^\text{Spin}_q=\sigma_q^\text{Spin}/\sigma_{DOS}^\text{Spin}$, with $h=0.5$ (circles) and $h=2$ (squares) as a function of $q$. Each point is an average over $10$ disorder realizations and a single initial state. Dotted lines are guides for the eyes. ${\cal{D}}=12,000$ for GOE and ${\cal{D}}=12,870$ ($L=16$) for the spin model.
    }
    \label{fig:sigma}
\end{figure}

For physical many-body quantum systems with two-body interactions, the density of states is Gaussian~\cite{French1970,Brody1981}. Thus, the expected shape of the LDoS for a system perturbed far from equilibrium and an initial state in the middle of the spectrum, as considered here, is also  Gaussian, as seen in Fig.~\ref{fig:LDoSspin1}~(c) for $q=2$ and $h=0.5$.

Despite the persistence of the Gaussian shape for different values of $q$, 
\begin{equation}
\label{eq:Spindos}
\rho_q^{\text{Spin}}(E)= \frac{1}{\sqrt{2\pi\sigma_q^2}}\exp\left[-\frac{(E_\alpha-E_{q}^{(0)})^2}{2\sigma_q^2}\right],   
\end{equation}
Fig.~\ref{fig:LDoSspin1} makes it clear that, in contrast to the GOE, the width $\sigma_q$ depends on $q$. As $q$ increases and the participation of the larger $|C_\alpha^{(0)}|^q$ gets amplified, the width of $\rho_q^{\text{Spin}}(E)$ gets narrower than the density of states. This indicates that the contributions from the components at the tails of the initial-state energy distribution, where chaotic states are nonexistent, get erased.

The dependence of the width of the gLDoS on $q$ reveals the limited degree of ergodicity of physical systems, even deep in the chaotic regime. Eigenstates of physical systems are not random vectors as in random matrices and are not random superpositions of plane waves as stated by the Berry's conjecture~\cite{Berry1977}. How to define chaotic states in realistic systems is discussed in Refs.~\cite{flambaum1994,flambaum1996a,flambaum1996b,flambaum1997,borgonovi1998,zelevinskyrep1996,borgonovi2016}. Our results add to these studies a way to quantify the level of ergodicity in comparison to random matrices.

In Fig.~\ref{fig:sigma}, we compare the results for $\sigma_q$ normalized by the width of the density of states (DoS) as a function of $q$ for the GOE model [Fig.~\ref{fig:sigma}~(a)] and the spin model [Fig.~\ref{fig:sigma}~(b)]. Each point in Fig.~\ref{fig:sigma} is obtained by performing an average over $10$ random realizations and a single initial state. The flat curve in Fig.~\ref{fig:sigma}~(a) indicates the presence of fully ergodic states throughout the spectrum, while in Fig.~\ref{fig:sigma}~(b) $\sigma_q^{Spin}/\sigma_{DOS}^{Spin}$ clearly decays as $q$ increases. This happens for the chaotic model with $h=0.5$ (circles), where nonchaotic states concentrate at the edges of the spectrum, and more abruptly for the case with $h=2$ (squares), where nonchaotic states are likely to be found also away from the edges of the spectrum.

\begin{figure}[htpb!]
    \centering
    \includegraphics[scale=0.4]{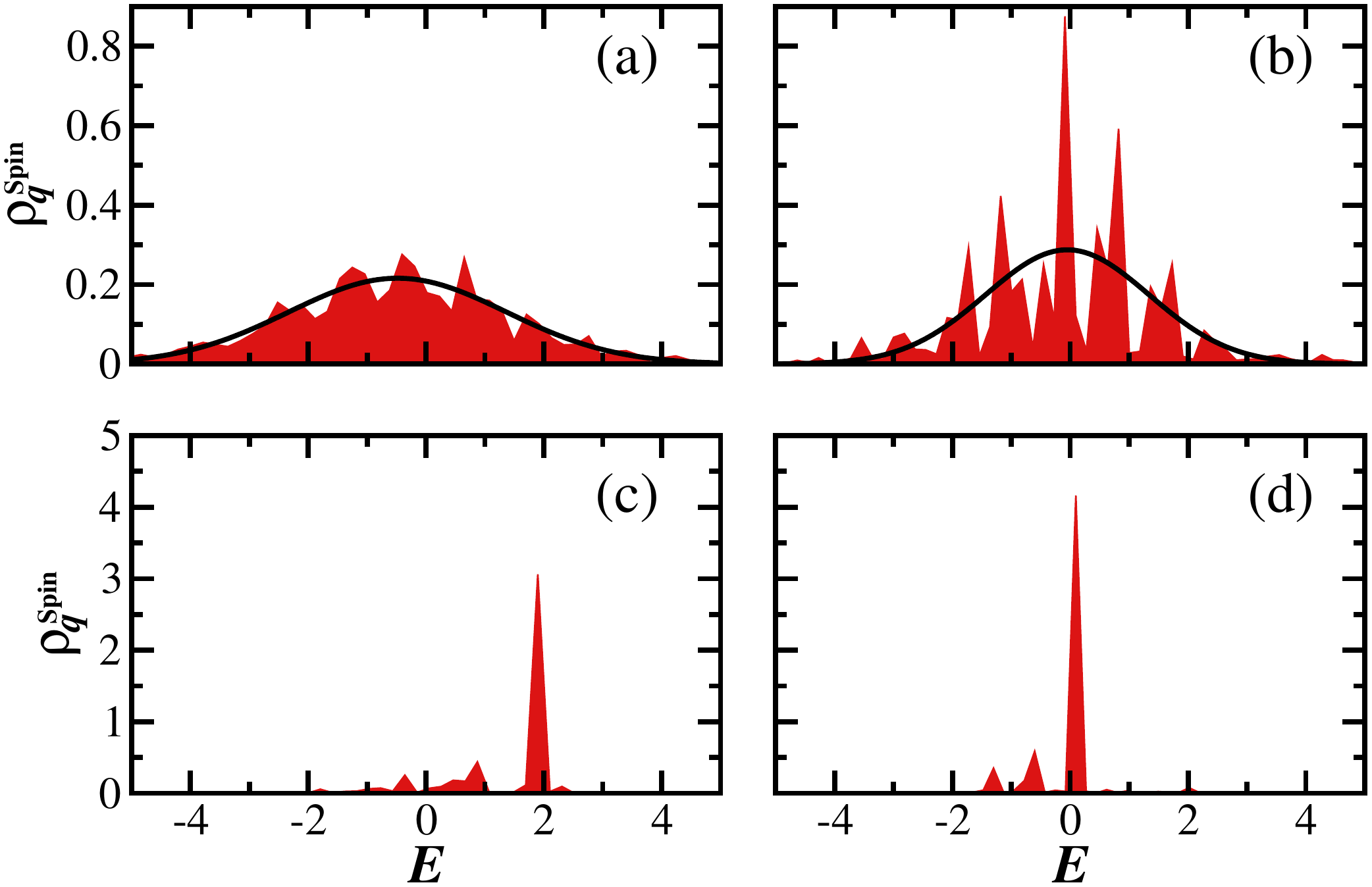}
    \caption{Generalized local density of states for the disordered spin-$1/2$ model with $h=2.0$ for (a) $q=0.5$, (b) $q=1.0$, (c) $q=2.0$, and (d) $q=3.0$. Shaded areas are numerical results and solid curves represent the Gaussian expression in Eq.~(\ref{eq:Spindos}). A single disorder realization and a single initial state are considered. The system size is $L=16$ with ${\cal{D}}=12,870$.
 }
    \label{fig:LDoSspin2}
\end{figure}

The reason for the abrupt decay of $\sigma_q^{Spin}$ with $q$ for $h=2$ becomes evident in Fig.~\ref{fig:LDoSspin2}, where we plot $\rho_q(E)$ for different values of $q$. When $q\leq1$ [Fig.~\ref{fig:LDoSspin2}~(a)-(b)], the shape of the LDoS is fragmented, while for $q>1$ [Fig.~\ref{fig:LDoSspin2}~(c)-(d)], this structure is nearly erased and $\rho_q(E)$ indicates a high degree of localization. 

For finite-size systems, several numerical studies supported that the eigenstates of the disordered spin model should become multifractal in its transition to the many-body localized phase~\cite{deluca2013,Luitz2014,li2015,Torres2015,Kohlert2019,Solorzano2021}, although this has not been confirmed in the thermodynamic limit~\cite{mace2019}. The patterns observed in Fig.~\ref{fig:LDoSspin2}~(a)-(b) also suggest fractality.

\section{Evolution of the generalized survival probability under the GOE model: Analytical expression}

\begin{figure}[htpb!]
    \centering
    \includegraphics[scale=0.3]{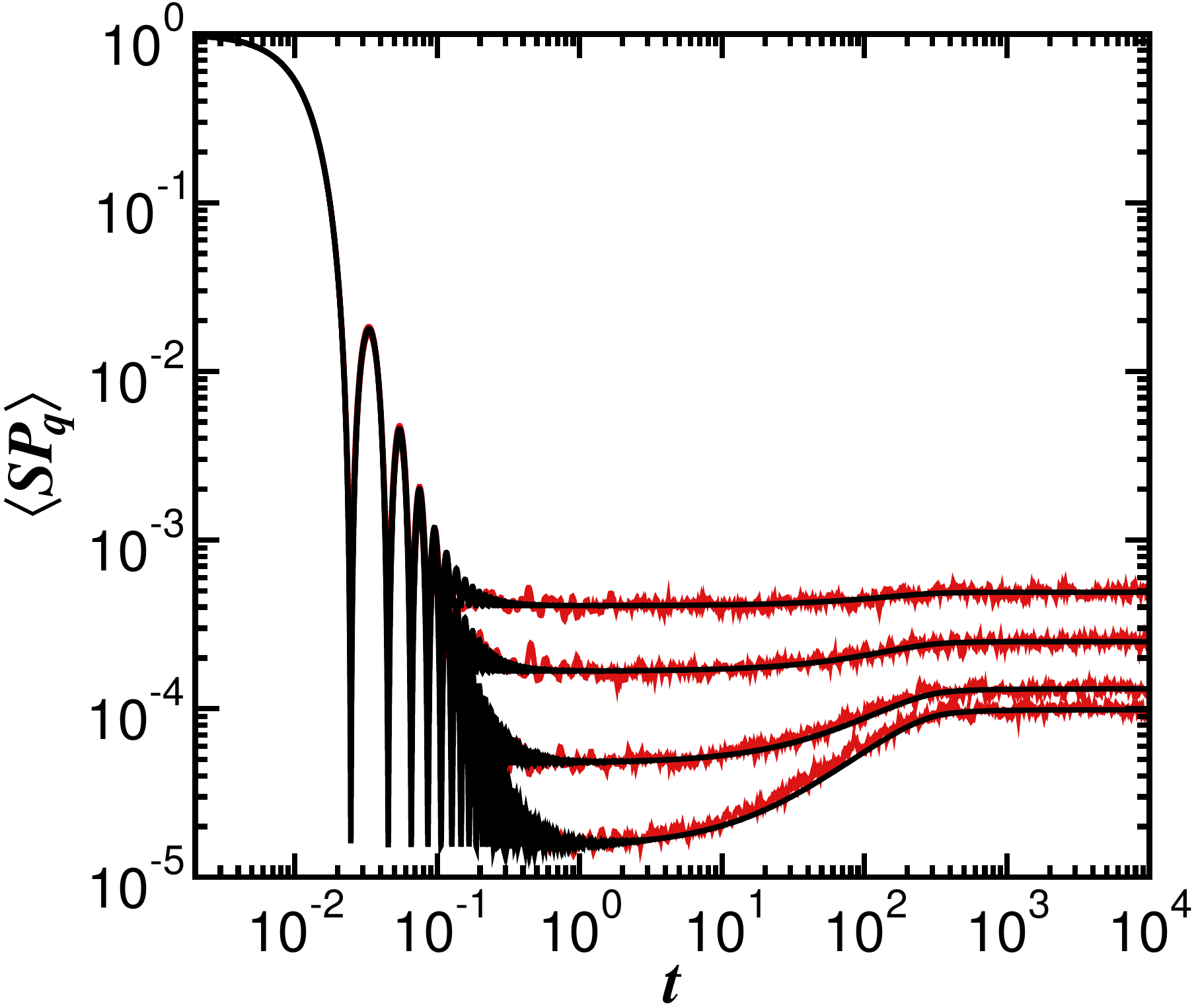}
    \caption{Generalized survival probability evolving under the GOE model for different values of $q$. Red curves are numerical results and the black lines correspond to the analytical expression in Eq.~\eqref{eq:sgpave}. From bottom to top, $q=0.5$, $1.0$, $2.0$ and $3.0$. Matrix size is ${\cal{D}}=12,000$. Averages over $10^4$ samples.}
    \label{fig:GSPGOE}
\end{figure}

According to Eq.~\eqref{eq:gsp}, the survival probability averaged over an ensemble of initial states and random realization is written as
\begin{equation}
\label{eq:gspa}
\left\langle SP_q(t)\right\rangle=\left\langle\frac{1}{{\cal{N}}_q^2}\sum_{\alpha\neq\beta}|C_\alpha^{(0)}|^q|C_\beta^{(0)}|^qe^{-i(E_\alpha-E_\beta)t}\right\rangle+\left\langle\frac{1}{{\cal{N}}_q^2}\sum_\alpha|C_\alpha^{(0)}|^{2q}\right\rangle,  
\end{equation}
where $\langle\dots\rangle$ denotes the average. The second term on the right-hand-side corresponds to the infinite-time average, $ \overline{SP}_q $, of the generalized survival probability. For GOE random matrices, where  $C_\alpha^{(0)}$ are random numbers from a  Gaussian distribution,
\begin{equation}
\label{eq:sat}
\overline{SP}_q = \frac{1}{{\cal{N}}_q^2}\sum_\alpha|C_\alpha^{(0)}|^{2q}  =\frac{\sqrt{\pi } \Gamma \left(q+\frac{1}{2}\right)}{{\cal{D}} \Gamma \left(\frac{q+1}{2}\right)^2},  
\end{equation}
Since for random matrices the eigenvalues and the eigenstates are statistically independent, they can be factorized (see details in~\cite{santostorres2017aip} and the appendix of~\cite{schiulaz2019}). Thus, using
\begin{equation}
\label{eq:eneave}
\left\langle e^{-i(E_\alpha-E_\beta)t}\right\rangle=\frac{1}{{\cal{D}}-1}\left[{\cal{D}}\frac{{\cal{J}}^2_1(2\sigma t)}{(\sigma t)^2}-b_2\left(\frac{\sigma t}{2{\cal{D}}}\right)\right],  
\end{equation}
and that
\begin{equation}
\label{eq:pref}
\frac{1}{{\cal{N}}_q^2}\sum_{\alpha\neq\beta}|C_\alpha^{(0)}|^q|C_\beta^{(0)}|^q=1-\overline{SP}_q,
\end{equation}
due to the requirement that $SP_q(t=0)=1$, 
we arrive at the analytical expression
\begin{equation}
\label{eq:sgpave}
\left\langle SP_q(t)\right\rangle=\frac{1-\left\langle\overline{SP}_q\right\rangle}{{\cal{D}}-1}\left[{\cal{D}}\frac{{\cal{J}}^2_1(2\sigma t)}{(\sigma t)^2}-b_2\left(\frac{\sigma t}{2{\cal{D}}}\right)\right]+\left\langle\overline{SP}_q\right\rangle.  
\end{equation}
Above, we wrote $\sigma_q=\sigma$, because $\sigma_q$ is constant for the GOE. The Fourier transform of the semicircular gLDoS gives the first term on the right-hand-side of Eq.~(\ref{eq:sgpave}), which involves the Bessel function of first kind, ${\cal{J}}_1$. This first term describes the initial decay of $\left\langle SP_q(t)\right\rangle$, as seen in Fig.~\ref{fig:GSPGOE}. It presents oscillations with the $nth$-zeros happening when the initial state dynamically finds out an orthogonal state at $t_n\sim (\pi  n+\sqrt{2}/2)/2 \sigma$ with $n=1,2,\dots$. The envelope of the oscillations decays as $t^{-3}$. The second term on the right-hand-side of Eq.~(\ref{eq:sgpave}), $b_2(t)=\left\{t \ln \left[(2 t+1)/(2 t-1)\right]-1\right\}\Theta (t-1) + [t \ln (2 t+1)-2 t+1]\Theta (1-t)$, is the so-called two-level form factor that takes $\left\langle SP_q (t)\right\rangle$ on a ramp to the saturation value $\left\langle\overline{SP}_q\right\rangle$.

In Fig.~\ref{fig:GSPGOE}, we compare the numerical results for $\left\langle SP_q(t)\right\rangle$ with the analytical expression in Eq.~\eqref{eq:sgpave}. The agreement is excellent. The fact that $\sigma_q$ for the GOE model is independent of $q$ (see Fig.~1 and Fig.~3) becomes evident once again in Fig.~\ref{fig:GSPGOE}, where the curves for different values of $q$ coincide at short times, capturing the oscillations of the Bessel function up to the minimum value of $\left\langle SP_q(t)\right\rangle$. 

To derive the time scale, $t^{GOE}_{Th}$, where $\left\langle SP_q(t)\right\rangle$ reaches the minimum of the correlation hole, we need to find the point where the first and second terms in the square brackets of Eq.~(\ref{eq:sgpave}) cross. Following Ref.~\cite{schiulaz2019}, we get the long-time expansion of the first term in Eq.~\eqref{eq:sgpave}
and expand the two-level form factor  for short times.
Combining the two in the derivative of $\left\langle SP_q(t)\right\rangle$ we get
\begin{equation}\label{eq:Thtime}
t^{GOE}_{Th} = \left( \frac{3}{\pi} \right)^{1/4} \frac{\sqrt{\cal{D}}}{\sigma} = \left( \frac{3}{\pi} \right)^{1/4}.
\end{equation}
To obtain the minimum value of $\left\langle SP_q(t)\right\rangle$ in the correlation hole, we evaluate Eq.~\eqref{eq:sgpave} at $t^{GOE}_{Th}$, which results in
\begin{equation}
\label{eq:spMin}
\begin{split}
\left.  \left\langle SP_q(t)\right\rangle  \right|_{t= t^{GOE}_{Th}} &\approx  \frac{1-\left\langle\overline{SP}_q\right\rangle}{{\cal{D}}-1}\left[ \frac{\cal{D}}{\pi (\sigma t^{GOE}_{Th} )^3} - \left( 1- \frac{\sigma}{\cal{D}} t^{GOE}_{Th}\right) \right] + \left\langle\overline{SP}_q\right\rangle\\
&\approx \frac{1-\left\langle\overline{SP}_q\right\rangle}{{\cal{D}}-1} (-1) + \left\langle\overline{SP}_q\right\rangle.
\end{split}
\end{equation}
Finally, using Eq.~\eqref{eq:sat} for $\left\langle\overline{SP}_q\right\rangle$ we arrive at
\begin{equation}
\label{eq:spMin2}
\left.  \left\langle SP_q(t)\right\rangle  \right|_{t= t^{GOE}_{Th}} \approx \frac{\sqrt{\pi } \Gamma \left(q+\frac{1}{2}\right)- \Gamma \left(\frac{q+1}{2}\right)^2}{{\cal{D}} \Gamma \left(\frac{q+1}{2}\right)^2}.   
\end{equation}
For the particular case of $q=2$, Eq.~\eqref{eq:spMin2} leads to the value $2/{\cal{D}}$ previously obtained in Ref.~\cite{schiulaz2019}. 

\section{Evolution of the generalized survival probability under the spin model}
\label{SPundeSPIN}

\begin{figure}[htpb!]
    \centering
    \includegraphics[scale=0.3]{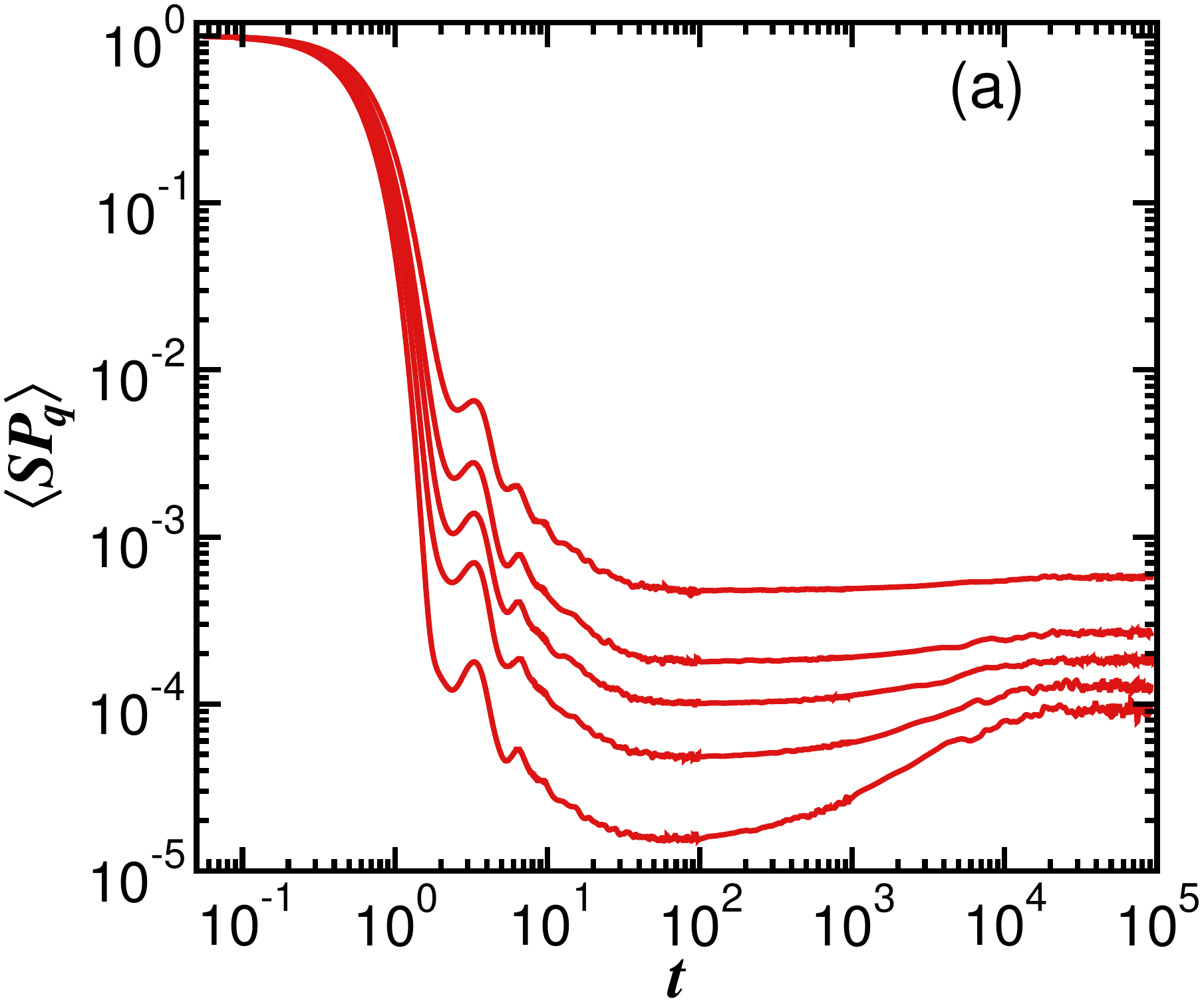}
    \includegraphics[scale=0.3]{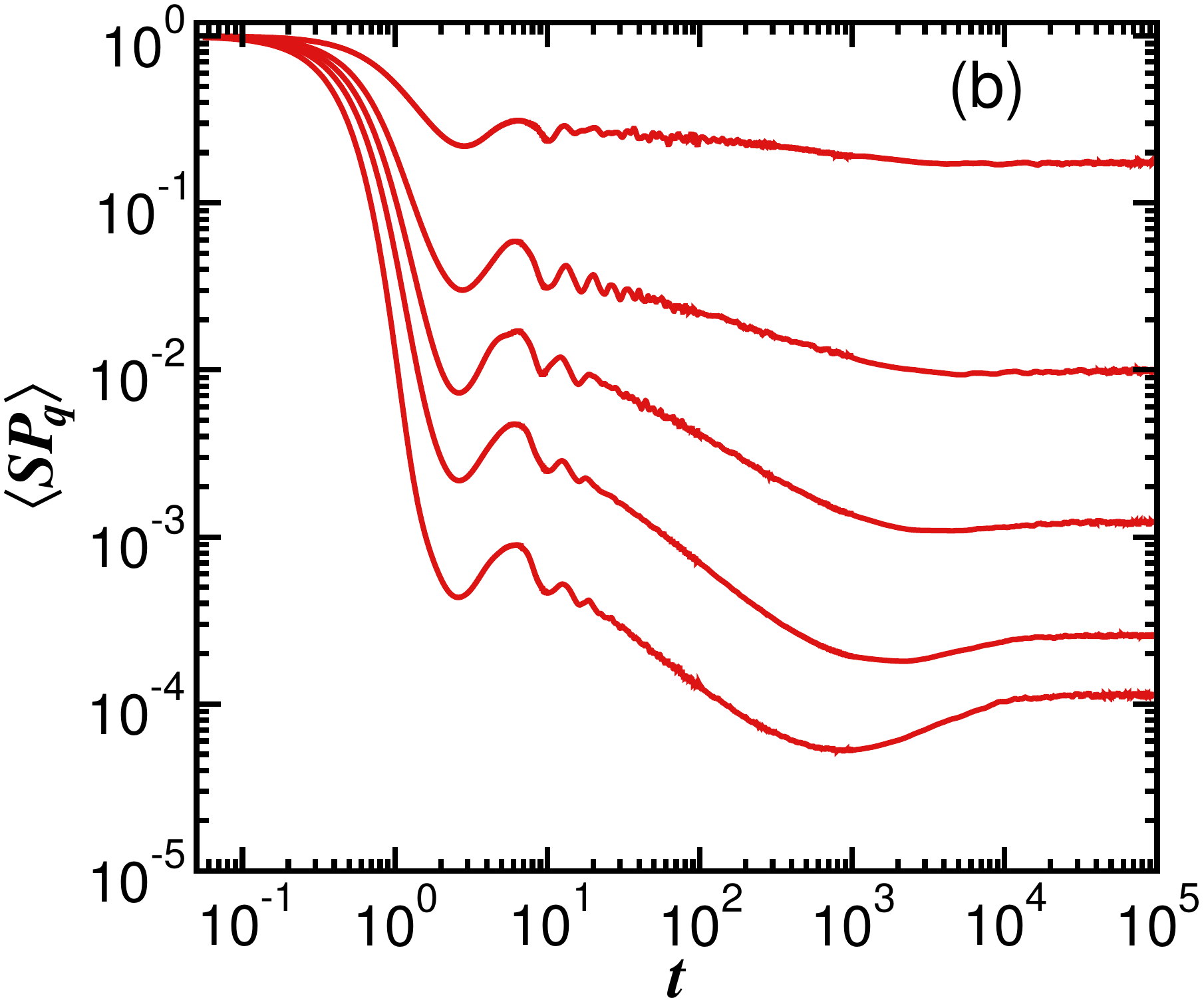}
    \caption{Generalized survival probability evolving under the disordered spin-$1/2$ model with (a) $h=0.5$ and (b) $h=2$ for different values of $q$. From bottom to top, $q=0.5$, $1.0$, $1.5$, $2.0$ and $3.0$. System size is $L=16$. Averages over $3\times 10^4$ samples.
    }
    \label{fig:GSPspin1}
\end{figure}

In Fig.~\ref{fig:GSPspin1}, we compare the evolution of $\left\langle SP_q(t)\right\rangle$ under the spin model for different values of $q$. In both panels, Fig.~\ref{fig:GSPspin1}~(a) for $h=0.5$ and Fig.~\ref{fig:GSPspin1}~(b) for $h=2$, the initial decay is determined by the envelope of the gLDoS, as seen from Eq.~\eqref{eq:gsp}. Since according to Fig.~\ref{fig:LDoSspin1} and Fig.~\ref{fig:LDoSspin2}, the shape of the distribution is Gaussian, one sees in Fig.~\ref{fig:GSPspin1} that 
\begin{equation}
\left\langle SP_q(t)\right\rangle = \exp(- \sigma_q^2 t^2)
\end{equation}
for $t\lesssim \sigma_q$. The dependence of $\sigma_q$ on $q$ is noticeable in Fig.~\ref{fig:GSPspin1}~(a) and evident in Fig.~\ref{fig:GSPspin1}~(b).

Beyond the Gaussian behavior, a power-law decay emerges,
\begin{equation}
\left\langle SP_q(t)\right\rangle \propto t^{-\gamma_q}.
\end{equation}
In Fig.~\ref{fig:GSPspin1}~(a), where the system is chaotic, the power-law exponent should depend on the bounds of the gLDoS. Since the gLDoS for the chaotic model in Fig.~\ref{fig:LDoSspin1} presents Gaussian tails for any $q$, we expect the same power-law exponent for all curves in Fig.~\ref{fig:GSPspin1}~(a), which is indeed what the nearly parallel lines during the algebraic behavior suggest. In contrast, in Fig.~\ref{fig:GSPspin1}~(b), it is clear that $\gamma_q$ decreases as $q$ increases and the minimum of the correlation hole takes longer to be reached. In this case, the power-law behavior reflects the correlations among the components of the initial state, which get enhanced for larger values of $q$.

\section{Discussion}

We introduced the concepts of generalized survival probability, $SP_q(t)$, and generalized local density of states, $\rho_q(E)$. We showed that the width of the generalized local density of states, $\sigma_q$, depends on $q$ even when the many-body quantum system is deep in the chaotic regime, which contrasts with random matrices, where the width is constant and equal to the width of the density of states. Therefore, $\sigma_q$ may serve as a tool to analyze and quantify the level of ergodicity of the states of physical systems with respect to random matrices. 

We also showed that the power-law behavior that follows the Gaussian decay of the generalized survival probability is strongly dependent on $q$ when the system is away from the chaotic regime. For a fixed value of the disorder strength, the power-law decay gets stretched as $q$ increases and the power-law exponent $\gamma_q$ decreases. This dependence of $\gamma_q$ on $q$ indicates correlations among the eigenstates. In a future work, we plan to investigate how $\sigma_q$ and $\gamma_q$ may be used to study multifractality.

\begin{acknowledgments}
D.A.Z.-H. and E.J.T.-H. have financial
support from VIEP-BUAP (Mexico), Project No. 00270. They are also grateful to LNS-BUAP for their supercomputing facility.  D.A.Z.-H. thanks CONACyT (Mexico) for financial support. L.F.S. is funded by the United States NSF Grant No. DMR-1936006. L.F.S. had support from the MPS Simons Foundation Award ID: 678586. E.J.T.-H. and L.F.S. are  grateful to Felix Izrailev for his constant support and invaluable teachings on quantum and classical chaos.
\end{acknowledgments}

\end{document}